\documentclass[aps,prl,twocolumn,showpacs,groupedaddress]{revtex4-1}  
\usepackage{graphicx}  
\usepackage{dcolumn}   
\usepackage{bm}        
\usepackage{amssymb}   

\begin{document}
\newcommand{\LNO}{$\rm{LiNbO_3}$\,}
\newcommand{\etal}{{\em et al.}}{}
\newcommand{\fig}[1]{Fig.~\ref{#1}}
\newcommand{\tab}[1]{{Table ~\ref{#1}}}
\newcommand{\AN}{$\rm{\AA}$\,}{}
\newcommand{\UPS}{$\rm{C/m^2}$\,\,}
\newcommand{\UPSS}{$\rm{\mu C/cm^2}$\,\,}
\newcommand{\UPY}{$\rm{C/m^2K}$\,\,}
\newcommand{\UPYY}{$\rm{\mu C/m^2K}$\,\,}
\newcommand{\ww}{0.5}

\title{Origin of Pyroelectricity in LiNbO$_3$}
\author{Qing Peng and R. E. Cohen}
\affiliation{
\begin{tabular}{cl}
 & Geophysical Laboratory, Carnegie Institution of Washington, 5251 Broad Branch Road NW, \\& Washington, D.C., 20015, U.S.A.\\
\end{tabular}
}

\begin{abstract}
We use molecular dynamics with a first-principles based shell model potential to study pyroelectricity in lithium niobate.
We find that the primary pyroelectric effect is dominant, and pyroelectricity can be understood simply from the anharmonic change in crystal structure with temperature and the Born effective charges on the ions.  This opens a new experimental route to studying pyroelectricity, as candidate pyroelectric materials can be studied with X-ray diffraction as a function of temperature in conjunction with theoretical effective charges. We also predict an appreciable pressure effect on pyroelectricity, which could be used to optimize materials pyroelectricity, and the converse electrocaloric effect, peak as $T_c$ is approached.
\end{abstract}


 \maketitle

The theory of ferroelectricity had a classical period that
culminated in the 1970s \cite{linesglass,PhysRevB.13.180, PhysRevLett.39.1362}, followed by a quiescent
period, and was rejuvenated in the 1990s with the introduction of
modern electronic structure methods to these complex, interesting,
and useful materials \cite{cohen:prb1990,cohen:Nature1992}. The fundamental physics of pyroelectricity, the change in polarization with respect to
temperature, has not been re-investigated until now, and there is no previous first principles computation of pyroelectricity. Pyroelectricity
is of current great interest since the discovery of particle
acceleration of ions from changes in temperature at
pyroelectric surfaces sufficient to generate hard X-rays in a commercial product
\cite{brownridge:640,ida2005,coolxx} as well as neutrons in heavy water via fusion
\cite{Naranjo:Nature2005}.   \LNO is a uniaxial pyroelectric with space group R3c in the polar
phase with ten atoms per primitive cell, and a $T_c$ of $1480$K
\cite{PhysRevB.53.1193,linesglass}. The structure, polarization and
lattice dynamics of \LNO have been previously studied from
first-principles using total energy, Berry's phase and linear
response methods within Density Functional Theory (DFT)
\cite{PhysRevB.53.1193,PhysRevB.61.272, PhysRevB.65.214302}. \LNO
has been studied extensively experimentally \cite
{Rauber,linesglass} due to its use in SAW filters and non-linear
optics. There is also much interest now in the converse of the pyroelectric effect, the electrocaloric effect, for refrigeration or energy scavenging. \cite{scott2007,Akcay2007,Prosandeev2008}

Pyroelectricity is the change in spontaneous polarization $P_s$ with
temperature $T$. The total pyroelectric coefficient is
\begin{equation}
\Pi = \frac{d P_s}{d T}=(\frac{\partial P_s}{\partial T})_\epsilon  + (\frac{\partial
P_s}{d\epsilon})_T  (\frac{\partial \epsilon}{\partial T})_\sigma
=\Pi_1+\Pi_2.
\label{totalPi}
\end{equation}
The first term on the right side is the primary pyroelectric effect and the second the
secondary effect.
Experimentally the pyroelectric effect is measured under the
constraint of constant stress. The experimentally accessible or proper pyroelectric
coefficient is due to the adiabatic current flow $J$ due to a
slow change in temperature, $\Pi'=\frac{dJ}{d\dot{T}}$, where
$\dot{T}$ is the change in temperature $T$ with time $t$. The $\Pi'$ of an unclamped sample can be expressed as
\begin{equation}
\Pi' = \Pi_1+\Pi_2+\Pi_3.
\label{equ:pyro}
\end{equation}
$\Pi_1$ measures the variation of
spontaneous polarization with respect to temperature at constant
strain (clamped), which arises from changes in phonon occupations and anharmonicity. $\Pi_2$ is the result of crystal deformation
where the strain caused by thermal expansion alters the polarization via the piezoelectric effect as 
${\Pi_2}_i = \alpha_{jk}c_{jklm}d_{ilm}$, where the indices label coordinate directions \cite{Newnham}, repeated indices imply summation, $d_{ilm}$ are
piezoelectric compliances, $c_{jklm}$ are elastic moduli, and
$\alpha_{ij}$ are the thermal expansion coefficients. $\Pi_3=2\alpha_1
P_s$ is the difference from the total and proper pyroelectric coefficients \cite{linesglass}, where $\alpha_1$ is the linear thermal expansion coefficient of the plane
perpendicular to the polar axis. $\Pi'$ can be measured with
charge-integration or dynamic pyroelectric
techniques\cite{linesglass}, whereas $\Pi_1,\Pi_2,\Pi_3$ cannot be measured directly. Understanding the components of $\Pi_1,\Pi_2,\Pi_3$ is crucial in studying pyroelectricity and its origin.

We performed density functional theory (DFT) computations and fitted the results to a atomistic shell model. 
We used Density Functional Perturbation Theory (DFPT) \cite{PhysRevB.43.7231} to compute phonons, effective charges, and dielectric constants.
We performed first principles calculations with the {\em ABINIT} package \cite{CMS.25.478} within the local density approximation (LDA) \cite{PhysRevB.45.13244}. Lithium $1s$ and $2s$ electrons, niobium
$4s$, $4p$, $4d$, and $5s$ electrons, as well as oxygen $2s $ and
$2p$ electrons were considered as valence states. We constructed pseudopotentials using the {\em
OPIUM} package\cite{opium}. 
We used a kinetic energy cutoff of 45 Hartree and sampled the Brillouin zone using a $6\times6\times6$ Monkhorst-Pack mesh of special $k$ points. The results were carefully checked against previous all-electron
\cite{PhysRevB.53.1193} and pseudopotential \cite{PhysRevB.61.272,PhysRevB.65.214302} computations.

\begin{table}
\caption{\label{tab:lc}First principles calculation of
 structure, spontaneous polarization $P_s$, constant volume specific heat capacity $\rm{C_v}$, volumetric thermal expansivity $\alpha$ of \LNO using DFPT and the ABINIT code.}
\begin{ruledtabular}
\begin{tabular}{c|c|c|c|c|c}
  &$a_H$&$c_H$&$P_s$&$\rm{C_v}$&$\alpha$\\
  & (\AN)&(\AN)&(\UPS)&($\rm{J/mol K}$)&($10^{-5}/K$)\\
\hline
DFT(0K)&5.151&13.703&0.86& &\\
\hline
QHLD(300K)&5.184&13.774&&94.04&3.59\\
\hline
MD(300K)&5.145&13.488&0.63&&2.63\\
\hline
Exp.(300K)&5.151$^a$&13.876$^a$&0.70-0.71$^b$&95.8$^c$&3.24-3.83$^d$\\
\end{tabular}
\end{ruledtabular}
$a$ \cite{Boysen:459};
$b$ \cite{PhysRevB.13.180,wemple:209};
$c$ \cite{LNO.cv};
$d$ \cite{LNO2002,kim:4637,smith:2219}
\end{table}

The structural parameters of \LNO in its polar ground
state as functions of volume were obtained by relaxing the cell shape and atomic positions
at seven volumes from $92.55$ to $110.19$ $\AN^3$(-5 to 20 GPa from
the fitted equation of state). 
We also optimized the atomic coordinates for the paraelectric symmetry
R$\bar{3}c$ with the same sets of optimized lattices.

We computed the phonon frequencies using DFPT
on a $4\times4\times4$ grid of q-points
at each of the seven volumes. The frequencies were interpolated onto
a finer grid using short-range force constants \cite{PhysRev.125.1905}. 
Quasi-harmonic Helmholtz free energies were
obtained from these frequencies as functions of temperature and
volume. Isotherms were fitted to the Vinet equation of
state \cite{PhysRevB.73.104303}. 

The polarization was computed for each of the seven volumes using the
Berry's phase method with a $4\times4\times20$ mesh of $k$
points. The results of the calculations were checked for convergence
with respect to the number of $k$ points and the plane wave cutoff
energy. We obtained $P_s$ as the differences of the polarization between
the polar and centrosymmetric \LNO at each volume, making sure we are on the same polarization versus mode coordinate curve \cite{resta2007}.

\begin{table}
\caption{\label{tab:piezo}First principles calculation of the elastic moduli $c$,
piezoelectric strain constants $d$ and piezoelectric stress constants $e$ of \LNO using DFPT and the ABINIT code.}
\begin{ruledtabular}
\begin{tabular}{cccc}
 & Smith \etal \cite{smith:2219} & Yamada \cite{amada:151} \etal & Present\\
\hline
$c$ & ($\times 10^{11}\rm{N/m^2}$) & ($\times 10^{11}\rm{N/m^2}$) & ($\times 10^{11}\rm{N/m^2}$)\\
$c_{11}$ &   2.030   &   2.03    &   2.18    \\
$c_{12}$ &   0.573   &   0.53    &   0.68    \\
$c_{13}$ &   0.752   &   0.75    &   0.78    \\
$c_{14}$ &   0.085   &   0.09    &   0.15    \\
$c_{33}$ &   2.424   &   2.45    &   2.40    \\
$c_{44}$ &   0.595   &   0.60    &   0.55    \\
$c_{66}$ &   0.728   &   0.75    &   0.75    \\
\hline
$d$ & ($\times 10^{-11}\rm{C/N}$) & ($\times 10^{-11}\rm{C/N}$) & ($\times 10^{-11}\rm{C/N}$)\\
$d_{15}$ &   6.92    &   6.8 &   8.12   \\
$d_{22}$ &   2.08    &   2.1 &   2.37    \\
$d_{31}$ &   -0.09   &   -0.1    &   -0.15   \\
$d_{33}$ &   0.60    &   0.6 &   0.81    \\
\hline
$e$ & ($\rm{C/m^2}$) & ($\rm{C/m^2}$) & ($\rm{C/m^2}$)\\
$e_{15}$ &   3.76    &   3.7 &   3.722    \\
$e_{22}$ &   2.43    &   2.5 &   2.317    \\
$e_{31}$ &   0.23    &   0.2 &   0.219    \\
$e_{33}$ &   1.33    &   1.3 &   1.718    \\
\end{tabular}
\end{ruledtabular}
\end{table}

\begin{figure}
\includegraphics[width=\ww\textwidth]{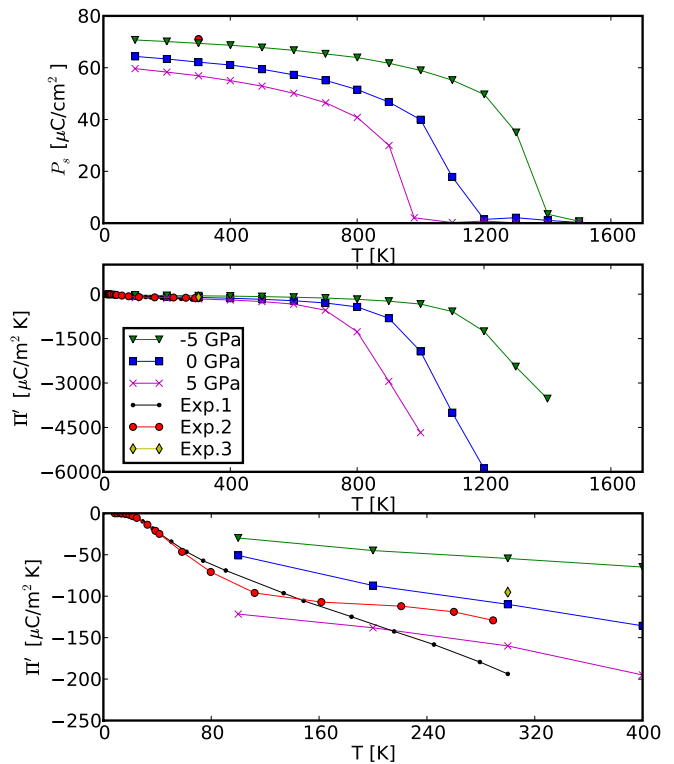}\\
\caption{\label{fig:NST} Molecular dynamics simulation results for \LNO in the
$N{\sigma}T$ ensemble. The results for $P=-5,0,5$ GPa are presented by triangle, square, and cross line respectively. Exp.1 is the experimental value of congruent sample in ref. \cite{PhysRevB.13.180}. Exp.2 is stoichiometric sample in ref. \cite{PhysRevB.13.180}; Exp.3 is experiment in ref. \cite{Lang:2005}. Note that our results from classical MD, so $\Pi'$ does not go to zero at zero temperature as required by quantum mechanics.}
\end{figure}

The shell model approach has proven to be a computationally
efficient and confident methodology for the simulation of
ferroelectric perovskites, including bulk properties of pure
crystals, solid solutions and super lattices, and also surfaces and
thin films properties\cite{Sepliarsky:2004}. In this model, each
atom is represented by a massive core coupled to a massless shell,
and the relative core-shell displacement describes the atomic
polarization. The model contains 4th
order core-shell couplings, long-range Ewald interactions and
short-range interactions described by the Rydberg potential $V(r) =
(a+br) \exp(r/\rho)$. The parameters were fit from the DFT and DFPT results of total energies, forces, stresses, phonon frequencies and eigenvectors, Born effective charges, and dielectric constants for a number of distorted and strained structures. We then performed classical molecular dynamic simulations with {\em DL\_POLY} package \cite{dlpoly}. 

We computed the spontaneous polarizations $P_s$ during the MD simulations. 
N$\sigma$T ensembles in MD simulations capture the evolutions of the system volume and shape corresponding to applied pressure, temperature. 
As a result,MD simulations in N$\sigma$T ensemble allow us to compute $dP_s/dT$, 
the total pyroelectric coefficient $\Pi$ in Eq. \ref{totalPi}. 
We also performed MD simulations in the N$\epsilon$T ensemble (constant strain) and obtained $\Pi_1$, and the difference gives $\Pi_2$. 
We computed $\Pi_3$ from MD N$\sigma$T simulations. 

The MD simulations were carried out in a super cell with $8 \times 8 \times 8$ primitive unit cells, which is 5120 atoms (5120 cores and 5120 shells). The N$\sigma$T ensemble allows the shape and volume to change at constant stress, by which $\Pi$ and $\Pi'$ can be obtained.  
$P_s$ decreases
with temperature and drops to zero at the phase transition to the paraelectric phase at 1200 K (\fig{fig:NST}), which agrees well with
the experimental value of 1430K \cite{Boysen:459} and 1480K \cite{linesglass}.

In order to understand the effects of volume error and the effects of pressure, we repeated the MD simulations and analysis at $\pm 5$ GPa. We found that $P=5$ GPa reduces the volume about 3.6\%, and $P=-5$ GPa increases the volume by 4.0\%.  $T_c$ is 1400 K and 1000 K for $P=-5$ and 5 GPa respectively.

\begin{figure}
\includegraphics[width=\ww\textwidth]{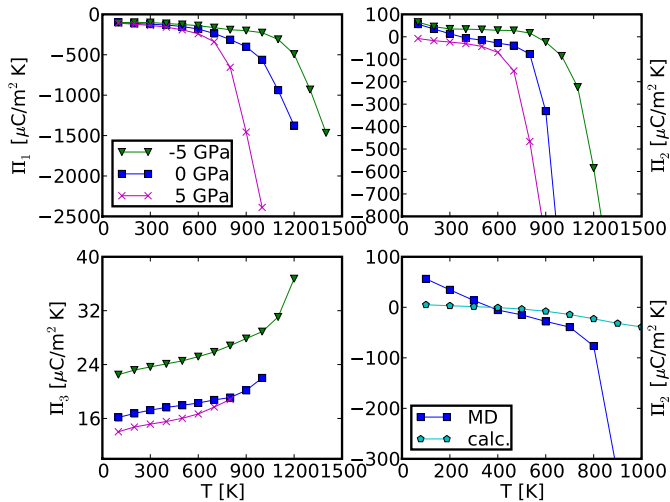}
\caption{\label{fig:pyro-MD} Pyroelectric coefficients. $\Pi_1$(top-left),$\Pi_2$(top-right),
$\Pi_3$(bottom-left) for $P=-5,0,5$ GPa, represented by triangle, square, cross line respectively.
Bottom-right panel shows $\Pi_2$ directly calculated from the MD (square line) simulations and computed with formula (pentagon line).}
\end{figure}

We separately computed $\Pi_1$ from MD simulations in the NVT ensemble.
The volume of the target temperature $T_v$
was taken from the previous $N\sigma T$ simulations.
MD simulations at $T=T_v,T_v\pm10,T_v\pm20$ K were carried out to calculate $\Pi_1$
at $T_v$ in \fig{fig:pyro-MD} and \tab{tab:pyro}. 
$\Pi_1$ decreases with temperature and pressures, as does $\Pi_2$, calculated by $\Pi-\Pi_1$.
We found $\Pi_2=13.5$ \UPYY and $\Pi_3=17.6$ \UPYY at 300 K. 
While lacking the direct and complete experimental data of all the coefficients of pyroelectricity, we estimate them as listed in in \tab{tab:pyro} by combining the reported data of ref.\cite{kim:4637,smith:2219} and \cite{PhysRevB.13.180}. There is good agreement between the experiment and present calculations.

\begin{table}
\caption{\label{tab:pyro}comparison of pyroelectric coefficient of \LNO.}
\begin{ruledtabular}
\begin{tabular}{c|ccccc}
  & $\Pi^{'}$ &$\Pi$ & $\Pi_1$ & $\Pi_2$ & $\Pi_3$ \\
\hline
Present & -90.2 & -107.7 &-121.3 & 13.5 & 17.6 \\
\hline
Calc. from Exp\cite{PhysRevB.13.180,smith:2219} & -133.0 &-154.9 &-171.9 & 17.0 & 21.9 \\
\hline
Exp \cite{Lang:2005} & &-83 & -95.8 & 12.8 &  \\
\end{tabular}
\end{ruledtabular}
\end{table}

As a check, we calculated $\Pi_2$ in an alternative way as 
$\Pi_2=\alpha_{jk}c_{jklm}d_{3lm}=2e_{31}\alpha_1+e_{33}\alpha_3$ for \LNO, where $e_{31},e_{33}$ are piezoelectric stress constants (Voigt notation), which are obtained by the first principles calculation at zero pressure and zero temperature as listed in Table \ref{tab:piezo}. Using $\alpha_j$ obtained from N$\sigma$T simulations, we computed $\Pi_2$ (\fig{fig:pyro-MD}), which agrees with direct MD results at low temperatures up to 700K.  

\begin{figure}
\includegraphics[width=\ww\textwidth]{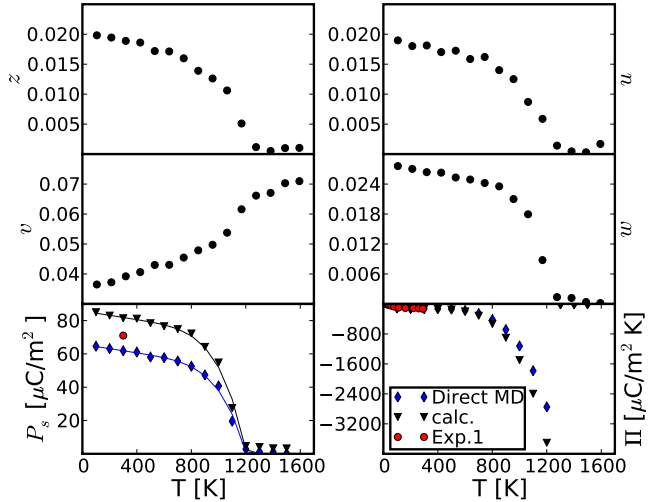}
\caption{\label{fig:bec} The average value of $z$, $u$, $v$ and $w$ in MD. The $P_s$ and $\Pi$ can be calculated using $z$, $u$, $v$, $w$ and Born effective charge $Z^*$ (triangle line), compared with experiment (circle line) \cite{PhysRevB.13.180} and direct MD results (diamond line).}
\end{figure}

We found that $\Pi_1$ is dominant and $\Pi_3$ is small. The absolute value of both $\Pi_1$ and $\Pi_2$ increase rapidly with temperature as $T_c$ is approached.

Up until now there has not been a clear exposition of the origin of pyroelectricity and the electrocaloric effect, but they are often considered as resulting from increasing polarization disorder with temperature. We find that the effects can be understood from the changes in crystal structure with temperature, as a simple anharmonic effect. We determined the average structural parameters $z$, $u$, $v$ and $w$ \cite{PhysRevB.65.214302} from the average atomic positions in the MD simulations (\fig{fig:bec}).  
We computed the $P_s$ versus temperature using these average positions with the Born effective charges $Z^*$ obtained from the DFPT computations, and 
$P_s=\frac{e}{\Omega} \sum_i Z^*_i r_i$ where
$r_i$ is the $i$th ionic displacement along the polar axis from the centrosymmetric to polar structures, $e$ the elementary charge and $\Omega$ the unit cell volume (\fig{fig:bec}).
The results show that the pyroelectric effect can be entirely understood in the classical regime above room temperature from the change in average structure with temperature, peaking at $T_c$.
Thus the anharmonic internal atomic rearrangement with respect to the temperature contribute the dominant part of the pyroelectricity. 
We find that pyroelectric coefficients could easily be obtained experimentally without electrical measurements, simply by studying changes in crystal structure with temperature, along with first-principles theoretical effective charges $Z^*$.
Good pyroelectrics and electrocaloric materials should have $T_c$ slightly higher than the operating temperatures.

This work was partly supported by the EFree, an Energy Frontier Research Center funded by the U.S. Department of Energy, Office of Science, Office of Basic Energy Sciences under Award Number DE-SC0001057 and partly by the Office of Naval Research No. N00014-07-1-0451.

\bibliography{pyro}

\end{document}